\documentclass{ws-procs11x85}
\usepackage{ws-procs-thm}           
\usepackage[table]{xcolor}
\definecolor{Gray}{gray}{0.9}
\usepackage{booktabs}
\definecolor{White}{rgb}{1,1,1}
\definecolor{Gray}{gray}{0.9}
\definecolor{LightCyan}{rgb}{0.88,1,1}
\usepackage{tabularx}
\usepackage{multirow}

\begin{document}

\title{Data-driven advice for applying machine learning to bioinformatics problems}

\author{Randal S.~Olson$^{\dag}$\footnote{\label{note1} Contributed equally}, William La Cava$^{\dag}\footnotemark[1]$, Zairah Mustahsan, Akshay Varik, and Jason H.~Moore$^{\dag}$}

\address{Institute for Biomedical Informatics, University of Pennsylvania\\
Philadelphia, PA 19104, USA\\
$^\dag$E-mails: rso@randalolson.com, lacava@upenn.edu and jhmoore@upenn.edu}

\begin{abstract}
As the bioinformatics field grows, it must keep pace not only with new data but with new algorithms. Here we contribute a thorough analysis of 13 state-of-the-art, commonly used machine learning algorithms on a set of 165 publicly available classification problems in order to provide data-driven algorithm recommendations to current researchers. We present a number of statistical and visual comparisons of algorithm performance and quantify the effect of model selection and algorithm tuning for each algorithm and dataset. The analysis culminates in the recommendation of five algorithms with hyperparameters that maximize classifier performance across the tested problems, as well as general guidelines for applying machine learning to supervised classification problems.
\end{abstract}


\copyrightinfo{\copyright\ 2018 Randal S.~Olson and William La Cava {\em et al}. Open Access chapter published by World Scientific Publishing Company and distributed under the terms of the Creative Commons Attribution Non-Commercial (CC BY-NC) 4.0 License.}
\bodymatter

\section{Introduction}

The bioinformatics field is increasingly relying on machine learning (ML) algorithms to conduct predictive analytics and gain greater insights into the complex biological processes of the human body~\cite{Bhaskar2006}. For example, ML algorithms have been applied to great success in GWAS, and have proven effective at detecting patterns of epistasis within the human genome~\cite{McKinney2006}. Recently, deep learning algorithms were used to detect cancer metastases on high-resolution pathology images~\cite{Liu2017Detecting} at levels comparable to human pathologists. These results, among others, indicate heavy interest in ML development and analysis for bioinformatics applications.

Owing to the development of open source ML packages and active research in the ML field, researchers can easily choose from dozens of ML algorithm implementations to build predictive models of complex data. Although having several readily-available ML algorithm implementations is advantageous to bioinformatics researchers seeking to move beyond simple statistics, many researchers experience ``choice overload'' and find difficulty in selecting the right ML algorithm for their problem at hand. As a result, some ML-oriented bioinformatics projects could be improved simply through the use of a better ML algorithm.

ML researchers are aware of the challenges that algorithm selection presents to ML practitioners. As a result, there have been some efforts to empirically assesses different algorithms across sets of problems, beginning in the mid 1990s with the StatLog project~\cite{king1995statlog}. Early work in this field also emphasized bioinformatics applications~\cite{tan2003empirical}. More recently, Caruana {\em et al.}~\cite{caruana2006empirical} and Fern\'{a}ndez-Delgado {\em et al.}~\cite{Delgado2014} analyzed several supervised learning algorithms, coupled with some parameter tuning. The aforementioned literature often compared many algorithms but on relatively few example problems (between 4 and 12), with only ~\cite{Delgado2014} using upwards of 112 example problems. In the time since these assessments, researchers have moved towards standardized, open source implementations of ML algorithms (e.g. scikit-learn~\cite{scikit-learn} and Weka~\cite{frank2004data}), and the number of publicly available datasets that can be used for comparison have skyrocketed, leading to the creation of decentralized, collaboration-based analyses such as the OpenML project~\cite{vanschoren2014openml}. However, the value of focused, reproducible ML experiments is still paramount. These observations motivated our work, in which we conduct a contemporary, open source, and thorough comparison of ML algorithms across a large set of publicly available problems, including several bioinformatics problems.

In this paper, we take a detailed look at 13 popular open source ML algorithms and analyze their performance across a set of 165 supervised classification problems in order to provide data-driven advice to practitioners who wish to apply ML to their datasets. A key part of this comparison is a full hyperparameter optimization of each algorithm. The results highlight the importance of selecting the right ML algorithm for each problem, which can improve prediction accuracy significantly on some problems. Further, we empirically quantify the effect of hyperparameter (i.e. algorithm parameter) tuning for each ML algorithm, demonstrating marked improvements in the predictive accuracy of nearly all ML algorithms. We show that the underlying behaviors of various ML algorithms cluster in terms of performance, as might be expected. Finally, based on the results of the experiments, we provide a refined set of recommendations for ML algorithms and parameters as a starting point for future researchers.

\section{Methods}

In this study, we compared 13 popular ML algorithms from scikit-learn\cite{scikit-learn}, a widely used ML library implemented in Python. Each algorithm and its hyperparameters are described in Table~\ref{tab:ml-params}. The algorithms include Na\"{i}ve Bayes algorithms, common linear classifiers, tree-based algorithms, distance-based classifiers, ensemble algorithms, and non-linear, kernel-based strategies. The goal was to represent the most common classes of algorithms used in literature, as well as recent state-of-the-art algorithms such as Gradient Tree Boosting~\cite{MachineLearningBook}.

For each algorithm, the hyperparameters were tuned using a fixed grid search with 10-fold cross-validation. In our results, we compare the average balanced accuracy~\cite{Velez2007} over the 10 folds in order to account for class imbalance. We used expert knowledge about the reasonable hyperparameters to specify the ranges of values to tune for each algorithm. It is worth noting that we did not attempt to control for the \textit{number} of total hyperparameter combinations budgeted to each algorithm. As a result, algorithms with more parameters have an advantage in the sense that they have more training attempts on each dataset. However, it is our goal to report as close to the best performance as possible for each algorithm on each dataset, and for this reason we chose to optimize each algorithm as thoroughly as possible.
 
The algorithms were compared on 165 supervised classification datasets from the Penn Machine Learning Benchmark (PMLB)~\cite{Olson2017PMLB}. PMLB is a collection of publicly available classification problems that have been standardized to the same format and collected in a central location with easy access via Python\footnote[1]{URL: \url{https://github.com/EpistasisLab/penn-ml-benchmarks}}. Although not limited to problems in biology and medicine, PMLB includes many biomedical classification problems, including tasks such as disease diagnosis, post-operative decision making, and exon boundary identification in DNA, among others. A sample of the biomedical classification tasks contained in PMLB is listed in Table~\ref{tbl:biomed_problems}.

Prior to evaluating each ML algorithm, we scaled the features of every dataset by subtracting the mean and scaling the features to unit variance. This scaling step was necessitated by some ML algorithms, such as the distance-based classifiers, which assume that the features of the datasets will be scaled appropriately beforehand.

The entire experimental design consisted of over 5.5 million ML algorithm and parameter evaluations in total, resulting in a rich set of data that is analyzed from several viewpoints in Section~\ref{s:results}. As an additional contribution of this work, we have provided the complete code required both to conduct the algorithm and hyperparameter optimization study, as well as access to the analysis and results\footnote[2]{URL: \url{https://github.com/rhiever/sklearn-benchmarks}}. Doing so allows researchers to easily compare algorithm performance on the datasets that are most similar to their own, and to conduct further analysis pertaining to their research.

\begin{table}

\centering

\tbl{ML algorithms and hyperparameters tuned in the experiments.}{\scriptsize
\label{tab:ml-params}
\begin{tabular}{l l}
\toprule
{\bf Algorithm} & {\bf Hyperparameters} \\
\midrule
Gaussian Na\"{i}ve Bayes (GNB) & No parameters.\\ \hline

\multirow{3}{*}{Bernoulli Na\"{i}ve Bayes (BNB)} & {\bf alpha}: Additive smoothing parameter.\\
 & {\bf binarize}: Threshold for binarizing the features.\\
 & {\bf fit\_prior}: Whether or not to learn class prior probabilities.\\ \midrule

\multirow{2}{*}{Multinomial Na\"{i}ve Bayes (MNB)} & {\bf alpha}: Additive smoothing parameter.\\
& {\bf fit\_prior}: Whether or not to learn class prior probabilities.\\ \midrule

\multirow{4}{*}{Logistic Regression (LR)} & {\bf C}: Regularization strength.\\
 & {\bf penalty}: Whether to use Lasso or Ridge regularization.\\
 & {\bf fit\_intercept}: Whether or not the intercept of the linear\\
 & classifier should be computed.\\ \midrule

\multirow{12}{*}{Stochastic Gradient Descent (SGD)} & {\bf loss}: Loss function to be optimized.\\
 & {\bf penalty}: Whether to use Lasso, Ridge, or ElasticNet\\
 & regularization.\\
 & {\bf alpha}: Regularization strength.\\
 & {\bf learning\_rate}: Shrinks the contribution of each successive \\
 & training update.\\
 & {\bf fit\_intercept}: Whether or not the intercept of the linear\\
 & classifier should be computed.\\
 & {\bf l1\_ratio}: Ratio of Lasso vs. Ridge reguarlization to use.\\
 & Only used when the `penalty' is ElasticNet.\\
 & {\bf eta0}: Initial learning rate.\\
 & {\bf power\_t}: Exponent for inverse scaling of the learning rate.\\ \midrule

\multirow{4}{*}{Passive Aggressive Classifier (PAC)} & {\bf loss}: Loss function to be optimized.\\
 & {\bf C}: Maximum step size for regularization.\\
 & {\bf fit\_intercept}: Whether or not the intercept of the linear\\
 & classifier should be computed.\\ \midrule

\multirow{5}{*}{Support Vector Classifier (SVC)} & {\bf kernel}: `linear', `poly', `sigmoid', or `rbf'.\\
 & {\bf C}: Penalty parameter for regularization.\\
 & {\bf gamma}: Kernel coef. for `rbf', `poly' \& `sigmoid' kernels.\\
 & {\bf degree}: Degree for the `poly' kernel.\\
 & {\bf coef0}: Independent term in the `poly' and `sigmoid' kernels.\\ \midrule

\multirow{2}{*}{K-Nearest Neighbor (KNN)} & {\bf n\_neighbors}: Number of neighbors to use.\\
& {\bf weights}: Function to weight the neighbors' votes.\\ \midrule

\multirow{6}{*}{Decision Tree (DT)} & {\bf min\_weight\_fraction\_leaf}: The minimum number of \\
 & (weighted) samples for a node to be considered a leaf.\\
 & Controls the depth and complexity of the decision tree.\\
 & {\bf max\_features}: Number of features to consider when\\
 & computing the best node split.\\
 & {\bf criterion}: Function used to measure the quality of a split.\\ \midrule

 & {\bf n\_estimators}: Number of decision trees in the ensemble.\\
Random Forest (RF) & {\bf min\_weight\_fraction\_leaf}: The minimum number of \\
 & (weighted) samples for a node to be considered a leaf.\\
\& & Controls the depth and complexity of the decision trees.\\
 & {\bf max\_features}: Number of features to consider when\\
Extra Trees Classifier (ERF) & computing the best node split.\\
 & {\bf criterion}: Function used to measure the quality of a split.\\ \midrule

\multirow{3}{*}{AdaBoost (AB)} & {\bf n\_estimators}: Number of decision trees in the ensemble. \\
 & {\bf learning\_rate}: Shrinks the contribution of each successive \\
 & decision tree in the ensemble. \\ \midrule

\multirow{8}{*}{Gradient Tree Boosting (GTB)} & {\bf n\_estimators}: Number of decision trees in the ensemble. \\
 & {\bf learning\_rate}: Shrinks the contribution of each successive \\
 & decision tree in the ensemble. \\
 & {\bf loss}: Loss function to be optimized via gradient boosting.\\
 & {\bf max\_depth}: Maximum depth of the decision trees.\\
 & Controls the complexity of the decision trees.\\
 & {\bf max\_features}: Number of features to consider when\\
 & computing the best node split.\\ \bottomrule
\end{tabular}}
\end{table}

\begin{table}
\tbl{A non-exhaustive sample of datasets included in the PMLB archive that pertain to biomedical classification.}
{\rowcolors{2}{white}{Gray}
\begin{tabular}{l r r r r} \toprule
Data Set & Classes & Samples & Dimensions & Description\\ \midrule
allbp	&	3	&	3772	&	29 & Diagnosis \\
allhyper	&	4	&	3771	&	29  & Diagnosis \\
allhypo	&	3	&	3770	&	29 & Diagnosis \\
ann-thyroid & 3 & 7200 & 21 &  Diagnosis  \\
biomed	&	2	&	209	&	8 & Diagnosis \\
breast-cancer-wisconsin	&	2	&	569	&	30 & Diagnosis \\
breast-cancer	&	2	&	286	&	9 & Diagnosis \\
diabetes	&	2	&	768	&	8 & Diagnosis \\
dna	&	3	&	3186	&	180 & Locating exon boundaries\\
GMT 2w-20a-0.1n&	2	&	1600	&	20 & Simulated GWAS\\
GMT 2w-1000a-0.4n	&	2	&	1600	&	1000 & Simulated GWAS\\
liver-disorder	&	2	&	345	&	6 & Diagnosis \\
molecular-biology\_promoters & 2 & 106 & 58 & Identify promoter sequences \\
postoperative-patient-data	&	2	&	88	&	8 & Choose post-operative treatment\\
\bottomrule
\end{tabular}}\label{tbl:biomed_problems}
\end{table}

\section{Results}
\label{s:results}

In this section, we analyze the algorithm performance results through several lenses. First we compare the performance of each algorithm across all datasets in terms of best balanced accuracy in Section~\ref{ap}. We then look at the effect of hyperparameter tuning and model selection in Section~\ref{tune}. Finally, we analyze how algorithms cluster across the tested problems, and present a set of algorithms that maximize performance across the datasets in Section~\ref{rec}.

\subsection{Algorithm Performance}\label{ap}

\begin{figure}
        \centering
        \includegraphics[width=\textwidth]{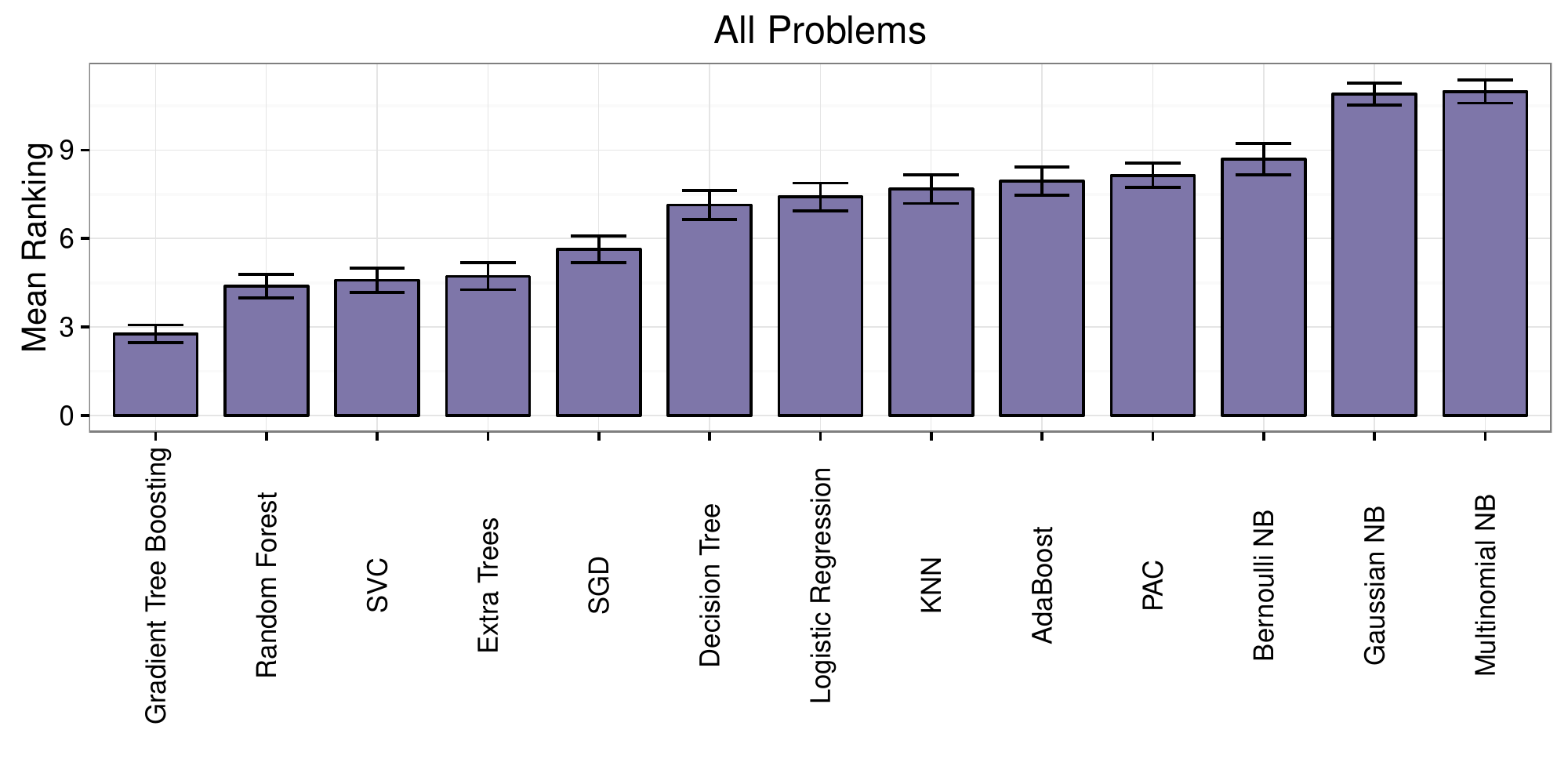}
        \caption{Average ranking of the ML algorithms over all datasets. Error bars indicate the 95\% confidence interval.}
        \label{fig:ranks}
\end{figure}

As a simple bulk measure to compare the performance of the 13 ML algorithms, we plot the mean rankings of the algorithms across all datasets in Figure~\ref{fig:ranks}. Ranking is determined by the 10-fold CV balanced accuracy of each algorithm on a given dataset, with a lower ranking indicating higher accuracy. The rankings show the strength of ensemble-based tree algorithms in generating accurate models: The first, second, and fourth-ranked algorithms belong to this class of algorithms. The three worst-ranked algorithms also belong to the same class of Na\"{i}ve Bayes algorithms.

In order to assess the statistical significance of the observed differences in algorithm performance across all problems, we use the non-parametric Friedman test~\cite{demsar2006statistical}. The complete set of experiments indicate statistically significant differences according to this test ($p < 2.2e^{-16}$), and so we present a pairwise post-hoc analysis in Table~\ref{tbl:fried}. The post-hoc test underlines the impressive performance of Gradient Tree Boosting, which significantly outperforms every algorithm except Random Forest at the $p<0.01$ level. At the other end of the spectrum, Multinomial NB is significantly outperformed by every algorithm except for Gaussian NB. These strong statistical results are interesting given the large set of problems and algorithms compared here. Because the No Free Lunch theorem~\cite{wolpert1997no} guarantees that all algorithms perform the same on average over all possible classes of problems, the differentiated results imply that the problems in the PMLB belong to a related subset of classes. The initial PMLB study~\cite{Olson2017PMLB} also noted the similarity in properties of several publicly available datasets, which could lead to inflated statistical significance. Nevertheless, it cannot be denied that the results are relevant to classification tasks encountered in real-world and biological contexts, since the vast majority of datasets used here are taken from those contexts.

\begin{figure}
    \centering
    \includegraphics[width=\textwidth]{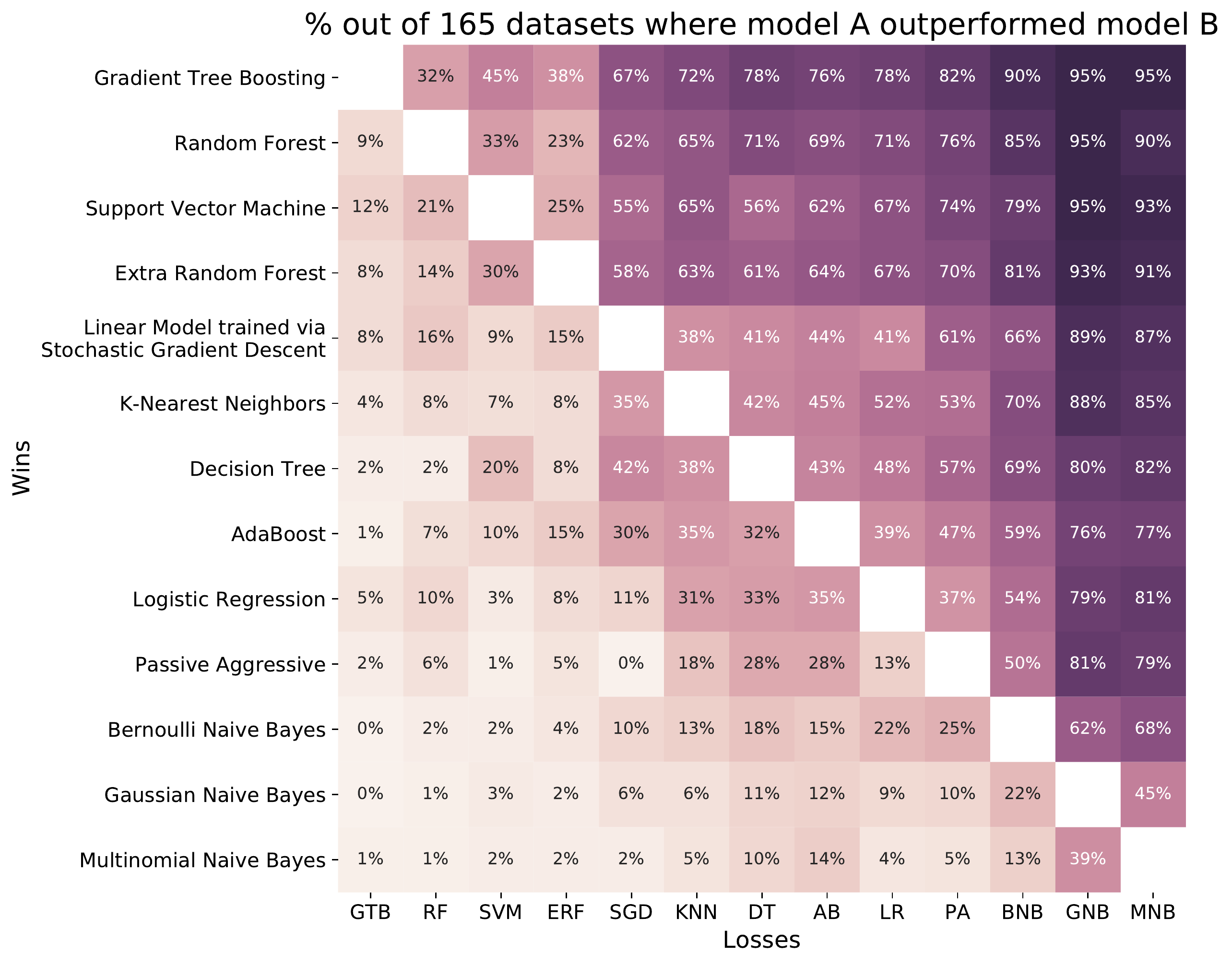}
    \caption{Heat map showing the percentage out of 165 datasets a given algorithm outperforms another algorithm in terms of best accuracy on a problem. The algorithms are ordered from top to bottom based on their overall performance on all problems. Two algorithms are considered to have the same performance on a problem if they achieved an accuracy within 1\% of each other.}
    \label{fig:outperform}
\end{figure}

Given these bulk results, it is tempting to recommend the top-ranked algorithm for all problems. However, this neglects the fact that the top-ranked algorithms may not outperform others for some problems. Furthermore, when simpler algorithms perform on par with a more complex one, it is often preferable to choose the simpler of the two. With this in mind, we investigate pair-wise ``outperformance'' by calculating the percentage of datasets for which one algorithm outperforms another, shown in Figure~\ref{fig:outperform}. One algorithm outperforms another on a dataset if it has at least a $1\%$ higher 10-fold CV balanced accuracy, which represents a minimal threshold for improvement in predictive accuracy.

In terms of ``outperformance,'' it is worth noting that no one ML algorithm performs best across all 165 datasets. For example, there are 9 datasets for which Multinomial NB performs as well as or better than Gradient Tree Boosting, despite being the overall worst- and best-ranked algorithms, respectively. Therefore, it is still important to consider different ML algorithms when applying ML to new datasets.

\begin{table}
\tbl{Post-hoc Friedman test of algorithm rankings across all problems. Bold values indicate $p<0.01$.}{\scriptsize
\rowcolors{2}{white}{Gray}
\begin{tabular}{l|rrrrrrrrrrrrr}
& GTB & RF & SVC & ERF & SGD & DT & LR & KNN & AB & PAC & BNB & GNB \\ \midrule
RF	&  { 0.01 }& -& -& -& -& -& -& -& -& -& -& -\\
SVC	& \textbf { 0.001 }&  { 1 }& -& -& -& -& -& -& -& -& -& -\\
ERF	& \textbf { 0.0004 }&  { 1 }&  { 1 }& -& -& -& -& -& -& -& -& -\\
SGD	& \textbf { 3e-10 }&  { 0.1 }&  { 0.4 }&  { 0.6 }& -& -& -& -& -& -& -& -\\
DT	& \textbf { 0 }& \textbf { 3e-09 }& \textbf { 1e-07 }& \textbf { 3e-07 }&  { 0.03 }& -& -& -& -& -& -& -\\
LR	& \textbf { 0 }& \textbf { 1e-11 }& \textbf { 1e-09 }& \textbf { 1e-07 }& \textbf { 0.003 }&  { 1 }& -& -& -& -& -& -\\
KNN	& \textbf { 0 }& \textbf { 1e-13 }& \textbf { 5e-12 }& \textbf { 7e-11 }& \textbf { 0.0002 }&  { 1 }&  { 1 }& -& -& -& -& -\\
AB	& \textbf { 0 }& \textbf { 6e-15 }& \textbf { 4e-14 }& \textbf { 4e-13 }& \textbf { 3e-06 }&  { 0.8 }&  { 1 }&  { 1 }& -& -& -& -\\
PAC	& \textbf { 0 }& \textbf { 2e-16 }& \textbf { 3e-15 }& \textbf { 8e-15 }& \textbf { 2e-07 }&  { 0.5 }&  { 0.9 }&  { 1 }&  { 1 }& -& -& -\\
BNB	& \textbf { 0 }& \textbf { 0 }& \textbf { 0 }& \textbf { 0 }& \textbf { 4e-10 }&  { 0.02 }&  { 0.1 }&  { 0.4 }&  { 0.9 }&  { 1 }& -& -\\
GNB	& \textbf { 0 }& \textbf { 0 }& \textbf { 0 }& \textbf { 0 }& \textbf { 0 }& \textbf { 0 }& \textbf { 2e-15 }& \textbf { 9e-13 }& \textbf { 1e-10 }& \textbf { 5e-09 }& \textbf { 2e-05 }& -\\
MNB	& \textbf { 0 }& \textbf { 0 }& \textbf { 0 }& \textbf { 0 }& \textbf { 0 }& \textbf { 0 }& \textbf { 2e-15 }& \textbf { 7e-14 }& \textbf { 1e-11 }& \textbf { 4e-09 }& \textbf { 4e-06 }&  { 1 }\\

\bottomrule
\end{tabular}}
\label{tbl:fried}
\end{table}

\subsection{Effect of Tuning and Model Selection}\label{tune}

Most ML algorithms contain several hyperparameters that can affect performance significantly (for example, the max tree depth of a decision tree classifier). Our experimental results allow us to measure the extent to which hyperparameter tuning via grid search improves each algorithm's performance compared to its baseline settings. We also measure the effect that model selection has on improving classifier performance.

\begin{figure}
        \centering
        \includegraphics[width=0.6\textwidth]{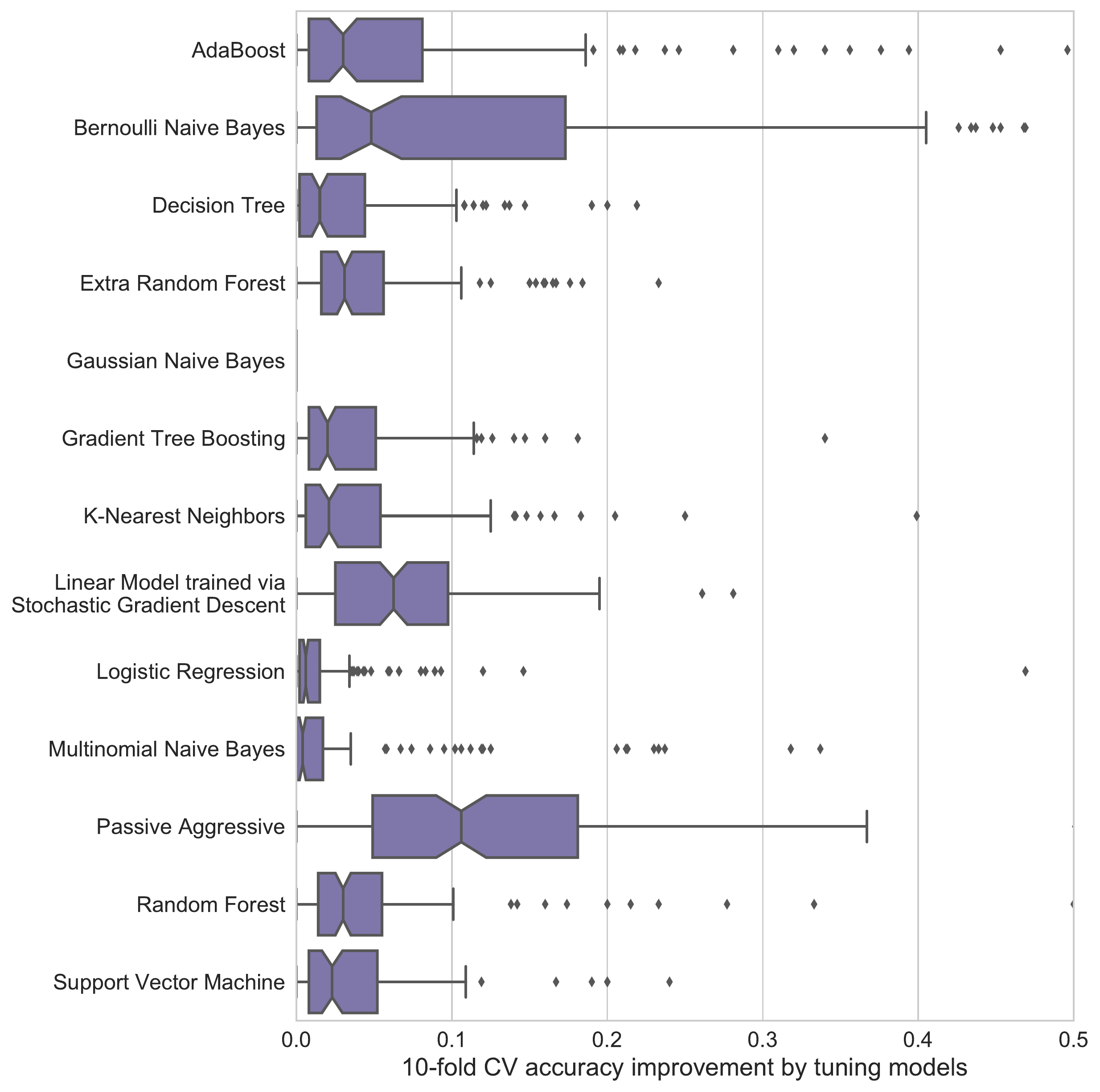}
        \caption{Improvement in 10-fold CV accuracy by tuning each ML algorithm's parameters instead of using the default parameters from scikit-learn.}
        \label{fig:tuned_acc}
\end{figure}

Figure~\ref{fig:tuned_acc} compares the performance of the tuned classifier to its default settings for each algorithm across all datasets. The results demonstrate why it is unwise to use default ML algorithm hyperparameters: tuning often improves an algorithm's accuracy by 3-5\%, depending on the algorithm. In some cases, parameter tuning led to CV accuracy improvements of 50\%.

\begin{figure}
        \centering
        \includegraphics[width=0.55\textwidth]{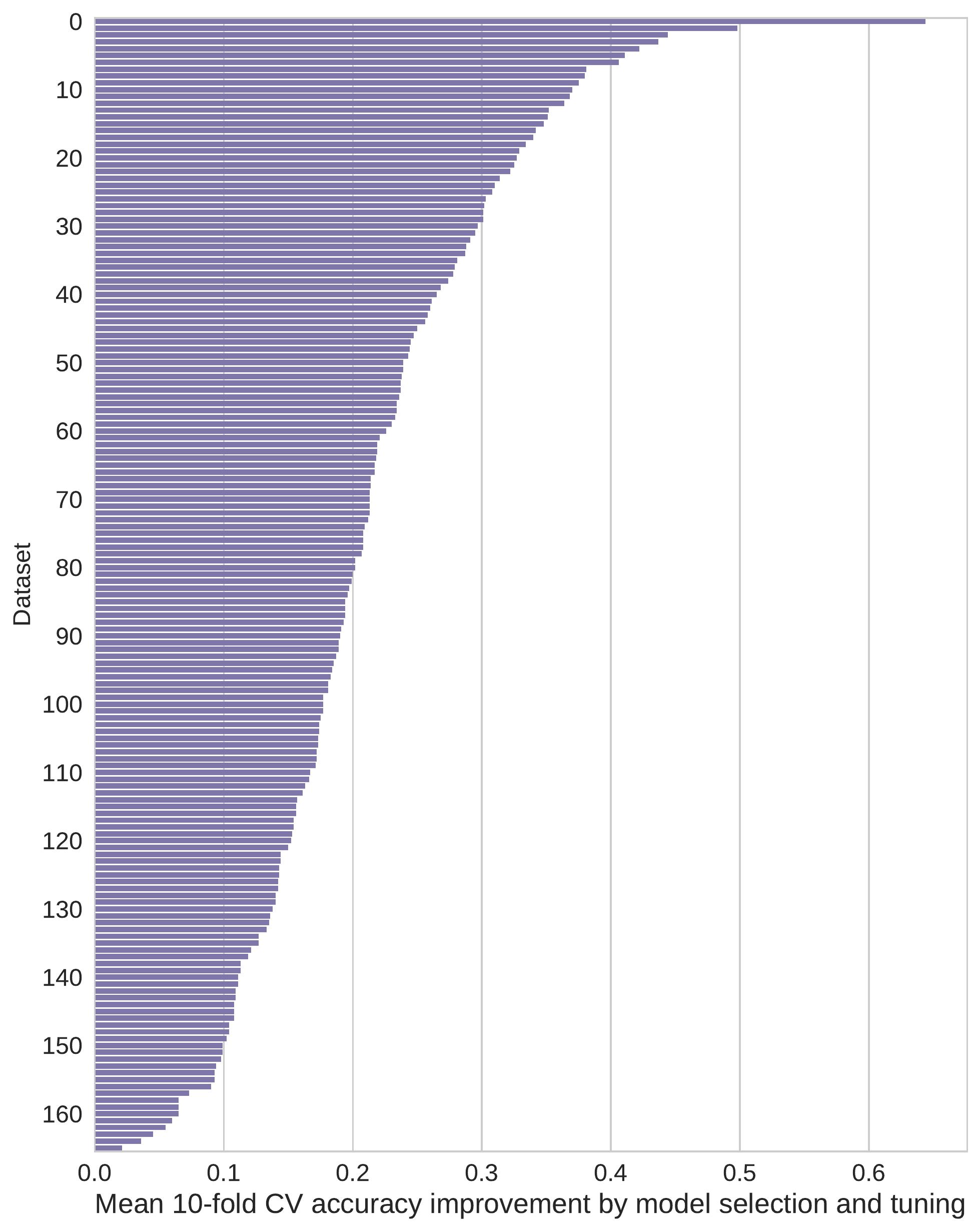}
        \caption{Improvement in 10-fold CV accuracy by model selection and tuning, relative to the average performance on each dataset.}
        \label{fig:tuned_all}
\end{figure}

Figure~\ref{fig:tuned_all} shows the improvement in 10-fold CV accuracy attained both by model selection and hyperparameter optimization compared to the average performance on each dataset. The results demonstrate that selecting the best model and tuning it leads to approximately a 20\% increase in accuracy, up to more than a 60\% improvement for certain datasets. Thus, both selecting the right ML algorithm and tuning its parameters is vitally important for most problems.

\subsection{Algorithm Coverage}\label{rec}

\begin{figure}
        \centering
        \includegraphics[width=0.6\textwidth]{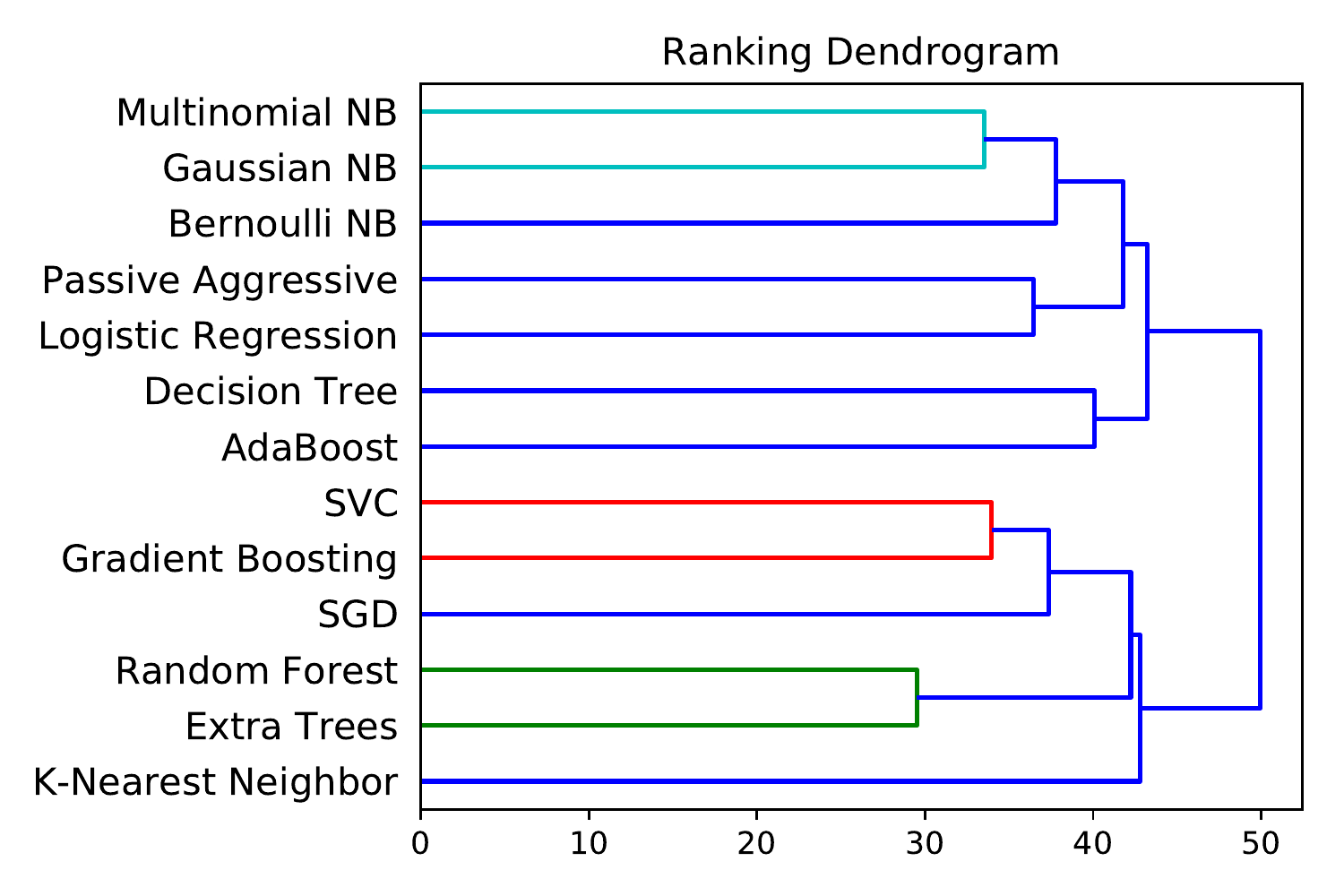}
        \caption{Hierarchical clustering of ML algorithms by accuracy rankings across datasets.}
        \label{fig:hac}
\end{figure}

Given that several of the 13 algorithms studied here have similar underlying methodologies, we would expect their performance across problems to align with the underlying assumptions that the modeling techniques have in common. One way to assess whether this holds is to cluster the performance of different algorithms across all datasets. We perform hierarchical agglomerative clustering on the 10-fold CV balanced accuracy results, which leads to the clusters shown in Figure~\ref{fig:hac}. Indeed, we find that algorithms with similar underlying assumptions or methodologies cluster in terms of their performance across the datasets. For example, the Na\"{i}ve Bayes algorithms (i.e., Multinomial, Gaussian, and Bernoulli) perform most similarly to each other, and the linear algorithms (i.e., passive aggressive and logistic regression) also cluster. The ensemble algorithms of Extra Trees and Random Forests, which both use ensembles of decision trees, also cluster. Support Vector Machines and Gradient Tree Boosting appear to be quite different algorithms, but given that both are able to capture nonlinear interactions between variables, it is less surprising that they cluster as well.

We present a list of five recommended algorithms and parameter settings in Table~\ref{tab:recommended-algorithms}. The five algorithms and parameters here are those that maximize the coverage of the 165 benchmark datasets, meaning that they perform within 1\% of the best 10-fold CV balanced accuracy obtained on the maximum number of datasets in the experiment. For the datasets in PMLB, these five algorithms and associated parameters cover 106 out of 165 datasets to within 1\% balanced accuracy. Notably, 163 out of 165 datasets can be covered by tuning the parameters of the five listed algorithms. Based on the available evidence, these recommended algorithms should be a good starting point for achieving reasonable predictive accuracy on a new dataset.

\begin{table}
\centering
\tbl{Five ML algorithms and parameters that maximize coverage of the 165 benchmark datasets. These algorithm and parameter names correspond to their scikit-learn implementations.}
{\begin{tabular}{|l|l|c|}
\hline
Algorithm & Parameters & Datasets Covered \\ \hline

\hline
GradientBoostingClassifier & loss=``deviance'' & \\
& learning\_rate=0.1 &  \\
& n\_estimators=500 & 51 \\
& max\_depth=3 & \\
& max\_features=``log2'' & \\ \hline

RandomForestClassifier & n\_estimators=500 & \\
& max\_features=0.25 & 19 \\
& criterion=``entropy'' & \\ \hline

SVC & C=0.01 & \\
& gamma=0.1 & \\
& kernel=``poly'' & 16 \\
& degree=3 & \\
& coef0=10.0 & \\ \hline

ExtraTreesClassifier & n\_estimators=1000 & \\
& max\_features=``log2'' & 12 \\
& criterion=``entropy'' & \\ \hline

LogisticRegression & C=1.5 & \\
& penalty=``l1'' & 8 \\
& fit\_intercept=True & \\ \hline
\end{tabular}}
\label{tab:recommended-algorithms}
\end{table}

\section{Discussion and Conclusions}

We have empirically assessed 13 supervised classification algorithms on a set of 165 supervised classification datasets in order to provide a contemporary set of recommendations to bioinformaticians who wish to apply ML algorithms to their data. The analysis demonstrates the strength of state-of-the-art, tree-based ensemble algorithms, while also showing the problem-dependent nature of ML algorithm performance. In addition, the analysis shows that selecting the right ML algorithm and thoroughly tuning its parameters can lead to a significant improvement in predictive accuracy on most problems, and is there a critical step in every ML application. We have made the full set of experiments and results available online to encourage bioinformaticians to easily gather information most pertinent to their area of study.

Even with a large set of results, it is difficult to recommend specific algorithms or parameter settings with a strong amount of generality. As a starting point, we provided recommendations for 5 different ML algorithms and parameters based on their collective coverage of the 165 datasets from PMLB. However, it is important to note that these algorithms and parameters will not work best on all supervised classification problems, and they should only be used as starting points. For a more nuanced approach, the similarity of the dataset on which ML is to be applied to datasets in PMLB could be quantified, and the set of algorithms that performed best on those similar datasets could be used. In lieu of detailed problem information, one could also use automated ML tools~\cite{olson2016evaluation,Feurer2015AutoSklearn} and AI-driven ML platforms~\cite{PennAI2017} to perform model selection and parameter tuning automatically.

Of course, some bioinformaticians may value properties of ML algorithms aside from predictive accuracy. For example, ML algorithms are often used as a ``microscope'' to model and better understand the complex biological systems from which the data was sampled. In this use case, bioinformaticians may value the interpretability of the ML model, in which case black box predictive models that cannot be interpreted are of little use~\cite{Ribeiro2016}. Although the logistic regression and decision tree algorithms are often outperformed by tree-based ensemble algorithms in terms of predictive accuracy (Figure~\ref{fig:outperform}), linear models and shallow decision trees often provide a useful trade-off between predictive accuracy and interpretability. Furthermore, methods such as LIME~\cite{Ribeiro2016} show promise for explaining why complex, black box models make individual predictions, which can also be useful for model interpretation.

There are several opportunities to extend the analysis in this paper in future work. A natural extension should be made to regression, which is used several biomedical applications such as quantitative trait genetics. In addition, these experiments do not take into account feature preprocessing, feature construction, and and feature selection, although it has been shown that learning better data representations can significantly improve ML performance~\cite{la2017ensemble}. We plan to extend this work to analyze the ability of various feature preprocessing, construction, and selection strategies to improve model performance. In addition, the experimental results contain rich information about the performance of different learning algorithms as a function of the datasets. In future work, we will take a deeper look into the properties of datasets that influence the performance of specific algorithms. By relating these dataset properties to specific areas of bioinformatics, we may be able to generate tailored recommendations for ML algorithms that work best for specific applications.

\section{Acknowledgments}

We thank Dr.~Andreas C.~M\"{u}ller for his valuable input during the development of this project, as well as the Penn Medicine Academic Computing Services for the use of their computing resources. This work was supported by NIH grants P30-ES013508, AI116794, DK112217 and LM012601, as well as the Warren Center for Network and Data Science at the University of Pennsylvania.

\bibliographystyle{ws-procs11x85}
\bibliography{references}

\end{document}